\documentclass[aps,prb,twocolumn,floatfix,showpacs,showkeys]{revtex4-1}
\usepackage[pdftex]{graphicx}
\usepackage{bm}
\usepackage{color}
\usepackage{amssymb}
\usepackage{mathrsfs}

\renewcommand{\deg}{$^{\circ}$\hspace{1mm}}
\newcommand{\etal}{ {\it et al.}\hspace{1mm}}
\newcommand{\newc}{\newcommand}
 
\newc{\be}{\begin{equation}}
\newc{\ee}{\end{equation}}
\newc{\bfe}{\begin{floatequation}}
\newc{\efe}{\end{floatequation}}
\newc{\bea}{\begin{eqnarray}}
\newc{\eea}{\end{eqnarray}}
\newc{\ie}{{\it i.e.} }
\newc{\eg}{{\it e.g.} }
\newc{\etc}{{\it etc.} }
\newc{\ra}{\rightarrow}
\newc{\lra}{\leftrightarrow}
\newc{\lsim}{\buildrel\langle\over{\sim}}
\newc{\gsim}{\buildrel\rangle\over{\sim}}

\newc{\one}{\mathbbm{1}}
\newc{\Tr}[1]{\mathrm{Tr}\left[ {#1} \right]}
\newc{\ket}[1]{\left|{#1}\right\rangle}
\newc{\bra}[1]{\left\langle{#1}\right|}
\newc{\braket}[2]{\langle{#1}|{#2}\rangle}
\newc{\mean}[1]{\langle{#1}\rangle}
\newc{\braketd}[1]{\langle{#1}|{#1}\rangle}
\newc{\ketbrad}[1]{\left|{#1}\rangle\!\langle{#1}\right|}
\newc{\ketbra}[2]{\left|{#1}\rangle\!\langle{#2}\right|}
\newc{\EV}[2]{\langle{#1}\rangle_{#2}}
\newc{\C}{\ensuremath{\mathbbm C}}

\def \bmlett{\begin{mathletters}}
\def \emlett{\end{mathletters}}
\def \r{{\bf r}}

\def \ra{\rightarrow}

\def\be{\begin{equation}}
\def\ee{\end{equation}}

\def\w01{\omega_{01}}
\def\r0{R_0}

\begin{document}

\title{Light production metrics of radiation sources}

\author{C. Tannous}
\affiliation{Laboratoire de Magn\'etisme de Bretagne - CNRS FRE 3117\\
UBO, 6, Avenue le Gorgeu C.S.93837 - 29238 Brest Cedex 3 - FRANCE}
\email{tannous@univ-brest.fr}
\date{\today}

\begin{abstract}
Light production by a radiation source is evaluated and reviewed
as an important concept of physics from the Black-Body point of view. 
The mechanical equivalent of the lumen, the unit of perceived light, 
is explained and evaluated using radiation physics arguments.
The existence of an upper limit of luminous efficacy is illustrated for
various sources and implications are highlighted. 
\end{abstract}


\maketitle

\section{Introduction}
Physics students are exposed to Black-Body radiation in undergraduate
Quantum Physics or in Graduate/Undergraduate Statistical Physics without
any clue regarding its significance as to the fundamental role it
played in the development of light source calibration, 
development and of lighting in general. 

Perhaps, only students with astrophysics or
atmospheric physics curriculum will be generally more aware of 
radiation physics in connection with the Black-Body fundamentals.

Presently, lighting is undergoing a tremendous evolution because of
the swift evolution of the light-emission-diode (LED) that is
now gaining larger and larger luminous efficacy to a point such that
it is now replacing, at a very impressive pace, our traditional home 
lighting, car headlights, LCD-monitors backlights, street lighting...

Additionally, LED colors are becoming more versatile and sharper
both in the case of traditional inorganic LED's or their organic
counterpart, the OLED. 

\begin{figure}[htbp]
  \centering
    \includegraphics[angle=0,width=80mm,clip=]{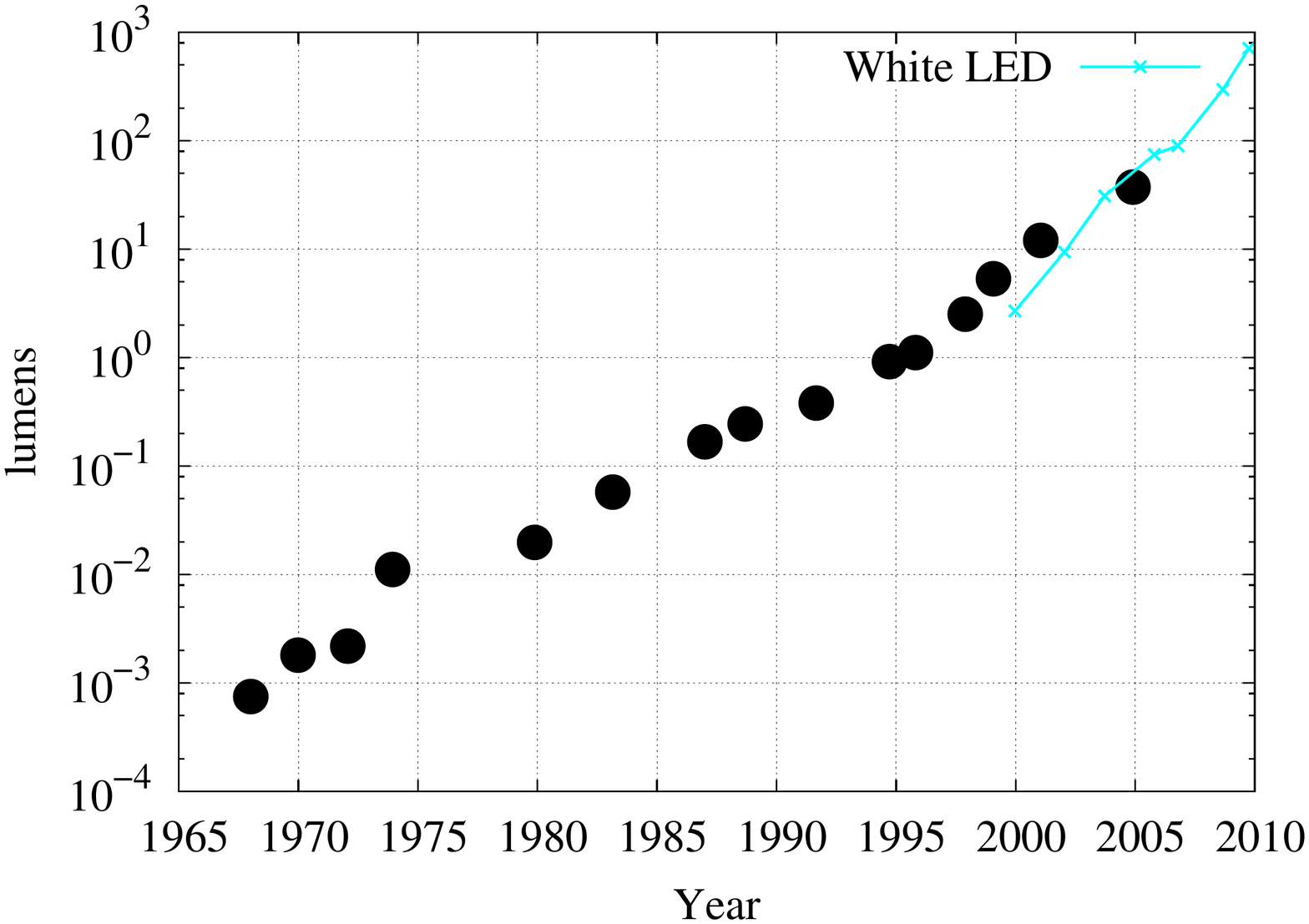}  
\caption{(Color online) Haitz rule and evolution of white LED (in cyan)
after year 2000 when S. Nakamura introduced In within GaN~\cite{PhysToday}.
Before 2000, evolution was driven by red LED starting in the 1970's with GaAs, GaAsP,
GaP, GaAsP:N, GaAlAs and finally GaAlInP. }
    \label{Haitz}
\end{figure}

The underlying basis of lighting progress is the existence
of Haitz rule (see fig.~\ref{Haitz}) that is similar to
Gordon Moore rule of Electronics evolution.

Light production by a radiation source is described by a luminous efficacy
ratio $\eta_L$ given by the product of $\eta_C$ the conversion
efficacy ratio of number of photons produced
(having any wavelength) to input energy (usually electrical 
but it could also be mechanical, thermal or chemical...) 
and $\eta_P$ the light perception efficacy ratio or Photometric 
efficacy ratio (PER). $\eta_P$ is the ratio of number of photons perceived 
by the human eye (photon wavelength in the visible spectrum) to
the total number of photons. Some authors call it 
$\eta_S$ the spectral efficacy.

This work can be taught as an application chapter 
in a general course of Statistical physics or in Semiconductor 
physics at the Undergraduate or Graduate level since 
physicists can contribute readily in improving light production
level of radiation sources or conversion efficacy
(through the increase of either $\eta_P$ or $\eta_C$) 
once the basis for luminous efficacy is
explained and illustrated along some notions of colorimetry and
light calibration.

The notions reviewed in this work are primarily concerned with 
$\eta_P$ the perception efficacy and are clearly important not 
only in Physics and technology but also for energy savings 
and efficiency, renewable energy and 
consequently for sustainable development of the Planet.

This report is organized as follows: In section 2, a review of lighting
metrology is made, in Section 3  luminous efficacy of radiation sources 
is explained and derived and in Section 4, we derive the maximum 
luminous efficacy on the basis of the colorimetry standard established by
the CIE~\cite{CIE}. This standard is briefly
explained and reviewed in the Appendix that comes after  
Section 5 carrying discussions and conclusions. 
 
\section{Metrology of lighting}
The SI system of units is based on seven entities: the meter, the kilogram,
the second, the ampere, the kelvin, the mole and the candela as the
unit of luminous intensity (Lumen is candela per unit solid angle).\\
Prior to 1979, the SI system of units defines the candela as follows: \\
{\it A pure sample of Platinum  at its fusion temperature (T=2042 K)
emits exactly 60 candelas/cm$^2$ along the vertical direction to the sample
and per unit solid angle}.

The SI system changed the definition during  
the 16th General Conference on Weights and Measures in October 1979 
to: \\
{\it The candela is the luminous intensity, in a given
direction, of a source that emits monochromatic radiation of frequency 
540 $\times 10^{12}$ Hz with a radiant intensity, in that direction,
of 1/683 Watts per unit solid angle}.

In order to relate these two definitions, despite the obvious
equivalence of 555 nm wavelength and 540 $\times 10^{12}$ Hz frequency,
some reminders about Black-Body radiation must be made.

Black-Body radiation was noticed for the first time by Gustav 
Kirchhoff~\cite{Kirchhoff,Grum}
who used to watch the color change of cavities present in heated metals 
worked by blacksmiths in his neighborhood. He noticed a systematic 
color change as the metal is being heated along the sequence of red, orange,
yellow, white and finally blue. 

Cavity color change did not depend on the nature of the metal but
depended on the size of the cavity.

This means that we have a photon gas in the cavity obeying  
Planck radiation law~\cite{Grum} with the average density of photons having  
energy $\hbar \omega$ of the Bose-Einstein form (the chemical potential
being zero since the number of photons is not fixed):

\be
n(\omega)= \frac{1}{ \left[ \exp(\frac{\hbar \omega}{k_B T}) -1 \right]}
\ee

The average radiation energy in the $[\omega, \omega+ d\omega]$ is
$(n(\omega)+\frac{1}{2})\hbar \omega  g(\omega)$ with $g(\omega)$ the 
density of states~\cite{density}. 
The vacuum energy $\frac{1}{2} \hbar \omega$ should not be included since it is
independent of $T$. The density of states~\cite{density} 
being $\frac{\omega^2}{\pi^2 c^3}$, we get:

\be
E(\omega)=  \frac{\omega^2}{\pi^2 c^3} \frac{\hbar \omega}{ \left[ \exp(\frac{\hbar \omega}{k_B T}) -1 \right]}
\ee

In order to get $E(\lambda)$ the average energy density of photons 
having wavelength $\lambda$, we use the conservation rule given by:

\be
E(\omega) d\omega= E(\lambda) d\lambda
\ee

with the transformation from angular frequency $\omega$ to linear
frequency $f$ to wavelength $\lambda$: $ \hbar \omega= \hbar 2 \pi f= h c/\lambda$.

This yields: 

\be
E(\lambda)= \frac{8 \pi h c}{\lambda^5} \frac{1}{ \left[ \exp(\frac{hc}{\lambda k_B T}) -1 \right]}
\ee

This law is derived in many Statistical Physics books and is usually enough for
standard Physics curriculum. It represents the thermal average number of photons per
unit volume emitted by a Black-Body at temperature $T$.

Since a Black-Body (absorbs and) emits radiation, we need a Planck
equivalent law for the emitted energy density. By analogy with electrical charge current
density ${\bm J}$ (${\bm J}= \rho {\bm v}$ with $\rho$ the charge density
and ${\bm v}$ their velocity), we multiply the 
average  photon density by the velocity of light $c$
and divide by $4\pi$, the whole space solid angle factor.
Thus we obtain the Planck distribution of photon energy emission at wavelength  
$\lambda$, temperature $T$ and unit solid angle:

\be
f_B(\lambda)= \frac{2 h c^2}{\lambda^5} \frac{1}{ \left[ \exp(\frac{hc}{\lambda k_B T}) -1 \right]}
\ee

Light sensed by the human eye is given by the distribution of energy average over
the eye sensitivity function called $V(\lambda)$ by the CIE~\cite{CIE}.

\begin{figure}[htbp]
  \centering
    \includegraphics[angle=0,width=80mm,clip=]{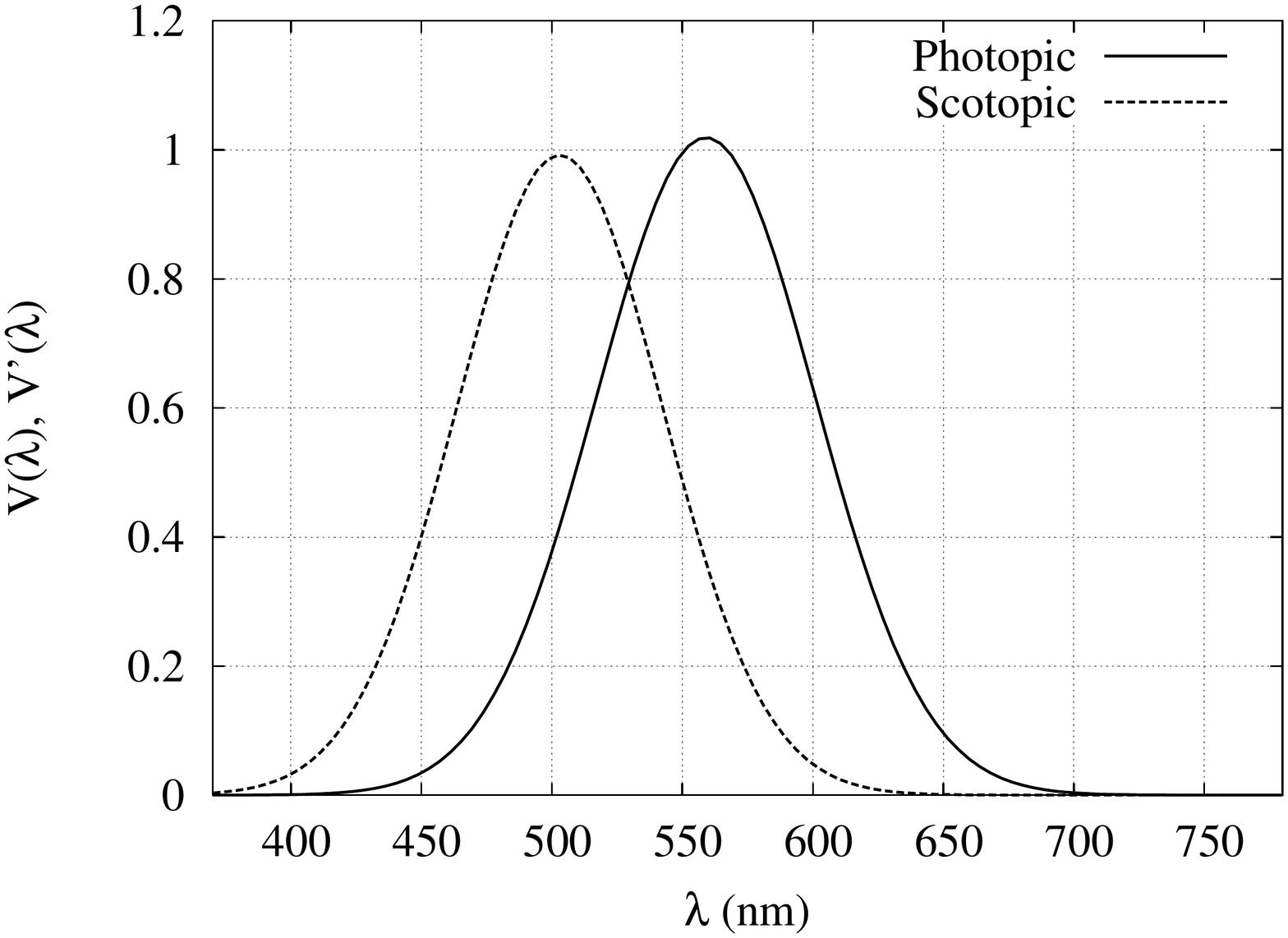}  
\caption{$V(\lambda)$ function giving the eye sensitivity versus wavelength in daylight
(photopic) and $V'(\lambda)$ in the dark (scotopic). They are shifted one with 
respect to the other by 45 nm, the Purkinje shift.}
    \label{vlam}
\end{figure}

This function depicted in fig.~\ref{vlam} peaks at a 555 nm (Yellow-Green) wavelength
that is the maximum sensitivity of the human eye. The analytical expressions of 
$V(\lambda)=1.019 \exp(-285 \left[(\frac{\lambda}{1000})-0.559\right]^2)$ (sensitivity in daylight or
photopic sensitivity) with $\lambda$ expressed in nm. The sensitivity in the dark $V'(\lambda)$ (scotopic) 
is shifted with respect to $V(\lambda)$ by 45 nm to shorter wavelengths and
peaks at 510 nm (Purkinje shift). Analytically 
$V'(\lambda)=0.992 \exp(-321.9 \left[(\frac{\lambda}{1000})-0.503\right]^2)$. 

The conversion from Radiometry $\lambda \in [0,\infty]$ to Photometry 
$\lambda \in [380 \mbox{nm}, 780 \mbox{nm}]$ is carried out by multiplying the 
total radiant power (in Watts) perceived by eye: 

\be
\int_0^\infty \frac{2 h c^2}{\lambda^5} \frac{1}{ \left[ \exp(\frac{hc}{\lambda k_B T}) -1 \right]} \times V(\lambda) d\lambda
\ee

by a conversion factor $K_m$ called the "Mechanical equivalent of the Lumen"
such that:

\be
\mbox{light perceived}= K_m \int_0^\infty \frac{2 h c^2}{\lambda^5} \frac{V(\lambda) d\lambda}{ \left[ \exp(\frac{hc}{\lambda k_B T}) -1 \right]}  
\ee

It is important to notice that we are carrying the integration over all positive
frequencies and not the visible spectrum given by the wavelength interval 
[380 nm, 780 nm] since we are dealing with (unlimited) radiation energy.

In order to evaluate $K_m$, we use the old definition of the candela and get:

\be
K_m= \frac{\mbox{60 cd/cm$^2$/sr}}{ \int_0^\infty \frac{2 h c^2}{\lambda^5} \frac{V(\lambda) d\lambda}{ \left[ \exp(\frac{hc}{\lambda k_B T}) -1 \right]} }
\ee

Transforming to SI units, the numerator becomes 6$\times 10^5$ lumens/m$^2$ 
(since lumen=cd per unit solid angle). The integral of the denominator
has Watt/m$^2$ dimensions and when it is numerically
evaluated we get $K_m=679$ lm/Watt (lumens/Watt) 
which is close to the value of 683 lm/Watt adopted by the SI~\cite{BIPM}.

\section{Luminous efficacy of radiation sources}

\subsection{Luminous efficacy of Black-Body radiation sources}
Having determined $K_m$ we are now in the position of determining the
luminous efficacy of any radiation source.

The value of $\eta_P$  for any radiation source 
characterized by a power emission spectrum  $P(\lambda)$ is given by:

\be
\eta_P=K_m \frac{ \int_{\mathscr{D}_\lambda} P(\lambda) V(\lambda) d\lambda}
{\int_{\mathscr{D}_\lambda} P(\lambda) d\lambda }
\label{PER}
\ee

The wavelength interval ${\mathscr{D}_\lambda}$ is arbitrary and depends on the radiation source.

In the particular case of a source of the thermal Black-Body type, we 
apply the above formula~\ref{PER} with $P(\lambda)=
f_B(\lambda)$ and ${\mathscr{D}_\lambda}= [0, \infty]$:

\be
\eta_P=K_m \frac{ \int_{0}^\infty f_B(\lambda) V(\lambda) d\lambda}
{ \int_{0}^\infty f_B(\lambda) d\lambda }
\label{PER_BB}
\ee

\begin{figure}[htbp]
  \centering
    \includegraphics[angle=0,width=80mm,clip=]{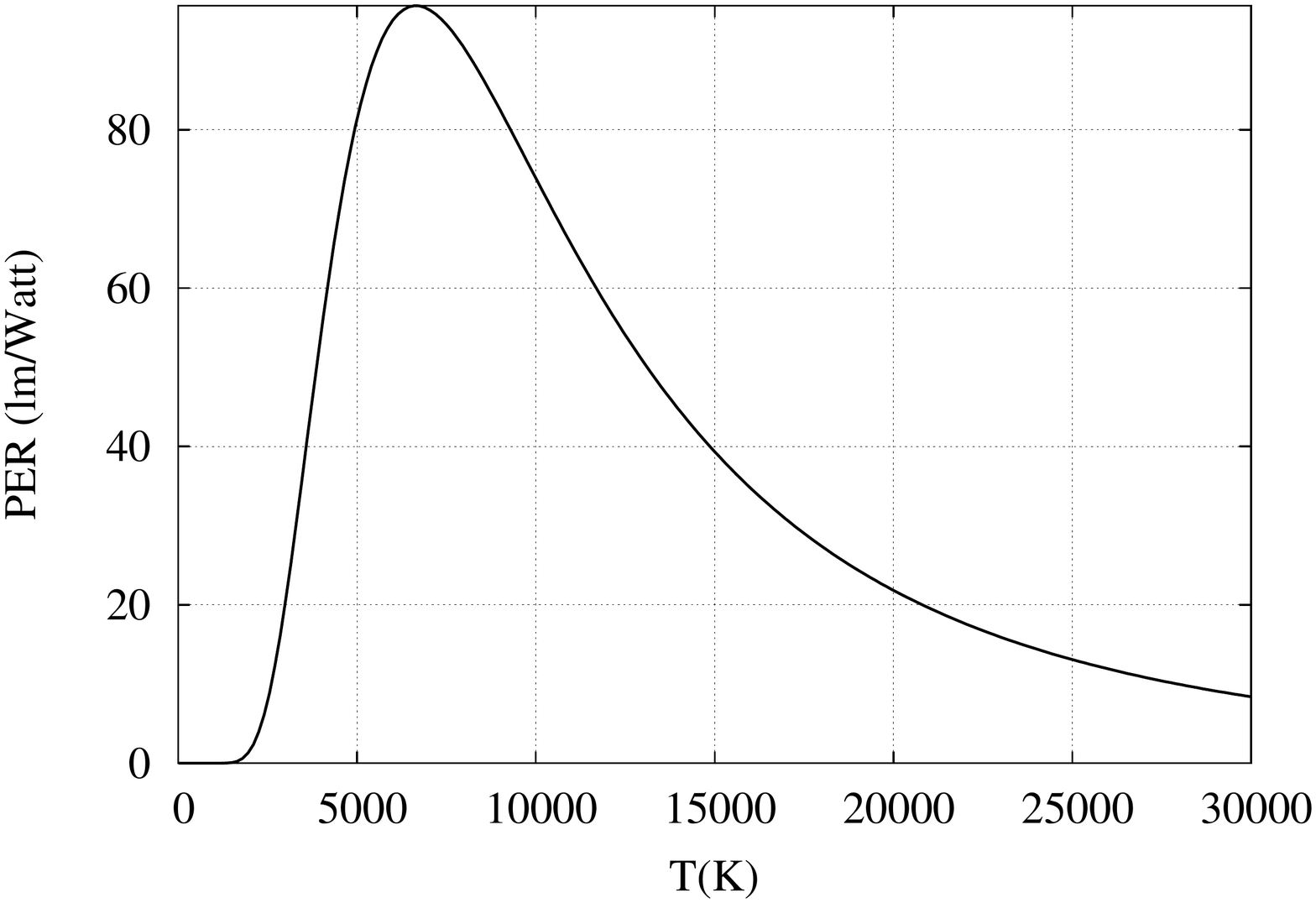}  
\caption{Photometric Efficacy Ratio $\eta_P$ versus temperature of the Black-Body. 
Notice that the maximum is 95 lm/Watt and that the temperature is about 7000K.}
    \label{PER_BBT}
\end{figure}

The result is the temperature dependent PER curve depicted in fig.~\ref{PER_BBT}.

PER has lm/Watt dimensions, thus we define {\it efficiency}
as a dimensionless ratio (in \%) yielding the fraction with respect to the ideal
efficacy of 683 lm/Watt.   

It shows that the Sun efficacy is about 93 lm/Watt (temperature about 6000K) 
or an efficiency of 93/683=13.6\%
and that the ordinary Tungsten light bulb based on incandescence 
phenomenom~\cite{Incand} is about 15 lm/Watt (temperature about 3000K)
or 2\% only. For a candle considered as a Black-Body at T=1800K, we get 0.6 lm/Watt
which corresponds to an efficiency of 0.6/683 or about 0.1\%.

The consequence in terms of lighting is that when one acquires a 60 Watts bulb
of the Tungsten type, the light produced by the bulb is 60 W x 15 lm/Watt= 900 lumens
in total and that has tremendous consequences for the quality and cost of the
lighting desired.

\subsection{Luminous efficacy of White radiation sources}
A White source is considered as having a flat power emission spectrum $P(\lambda)$ over
the entire visible interval,
nevertheless in practice the interval is limited and one has to define precisely the 
wavelength interval over which this flatness is observed.

Two cases are encountered in lighting systems:

\begin{enumerate}

\item White source as a truncated Black-Body source: \\
This is a Black-Body source taken at a temperature T=5800K with 
a spectrum limited by definition to $\lambda_{min}$=400 nm and 
$\lambda_{max}$=700 nm. The PER is obtained from:

\be
\eta_P=K_m \frac{  \int_{\lambda_{min}}^{\lambda_{max}} P(\lambda) V(\lambda) d\lambda}
{ \int_{\lambda_{min}}^{\lambda_{max}} P(\lambda) d\lambda }
\label{PER_TB}
\ee

Using $P(\lambda)= \frac{2 h c^2}{\lambda^5} \frac{1}{ \left[ \exp(\frac{hc}{\lambda k_B T}) -1 \right]}$ we get a PER of about 250 lm/Watt.

\item Equal Energy White Source: \\
For instance, an "Equal Energy White Source" possesses by definition a flat 
power emission spectrum over the entire visible spectrum. Mathematically $P(\lambda)=W$
for $\lambda \in [380 \mbox{nm}, 780 \mbox{nm}]$ i.e. $\lambda_{min}$=380 nm, whereas 
$\lambda_{max}$=780 nm. Thus we get:

\be
\eta_P=K_m \frac{\int_{\lambda_{min}}^{\lambda_{max}} W V(\lambda) d\lambda}
{\int_{\lambda_{min}}^{\lambda_{max}} W d\lambda }
\label{PER_EEW}
\ee

This yields about 179 lm/Watt. This leads to the conclusion that flatness is not
enough to increase PER. We need a compromise between flatness and wavelength 
interval length.

\subsection{Luminous efficacy of fluorescent sources}
Fluorescent light sources are known as cold sources (Eco light is also 
a fluorescent light source) as opposed to thermal (Black-Body
like) or incandescent sources. They need a special circuit called a ballast to
stabilize current and accelerate electrons in order to make them 
collide inelastically with a gas mixture of heavy atoms (typically
Mercury, Terbium and Argon) producing radiation.

An example power emission spectrum of the three band type is displayed in fig.~\ref{fluo}.
It shows several peaks over a finite wavelength interval 
in sharp contrast with Black-Body spectrum that is smooth and
continuous extending over an infinite wavelength interval.

\begin{figure}[htbp]
  \centering
    \includegraphics[angle=0,width=80mm,clip=]{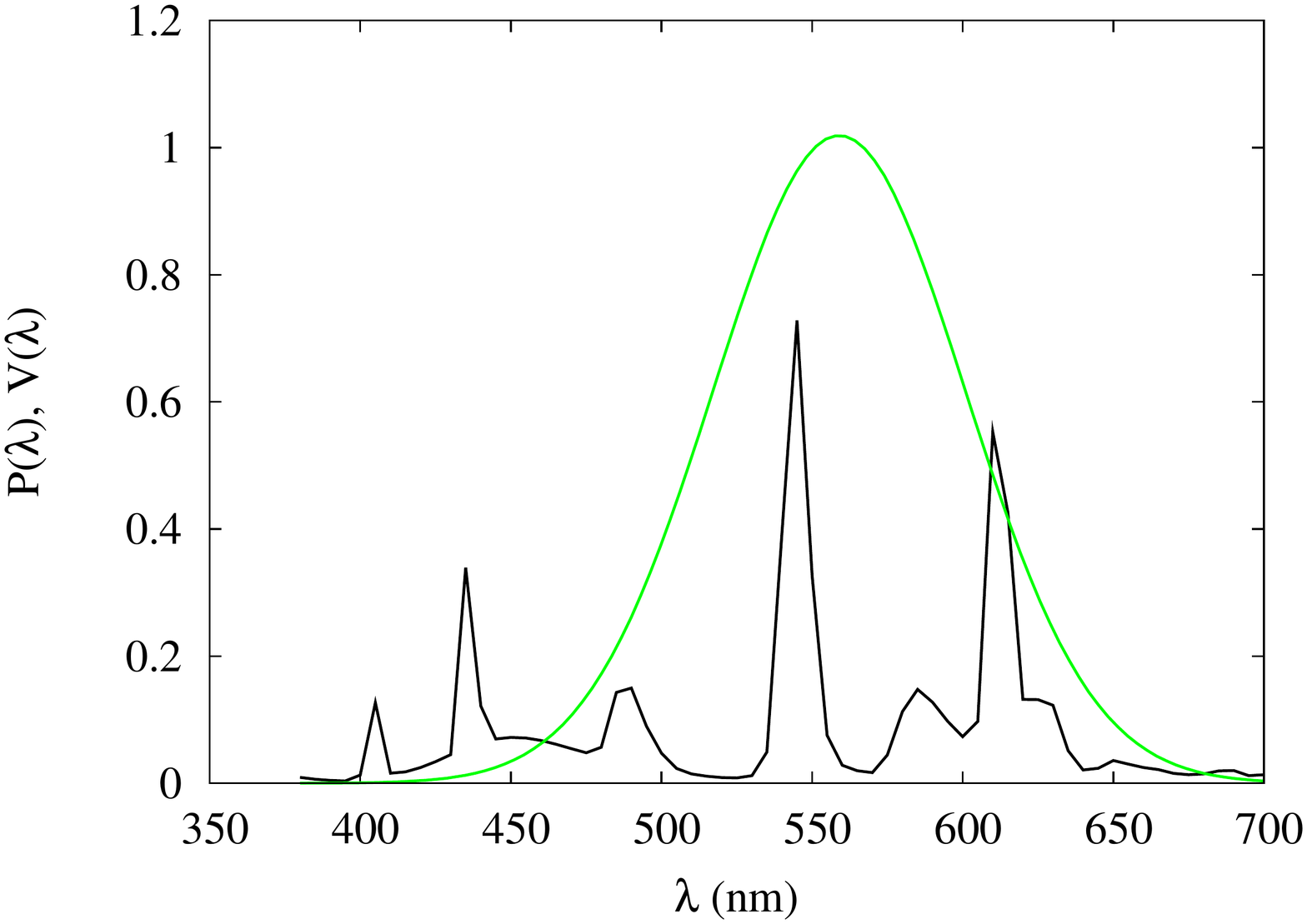}  
\caption{(Color online) Relative power emission spectrum of a three-band 
type fluorescent bulb compared to eye sensitivity curve (in green). 
Mercury peak is around 450 nm (Data adapted from Hoffmann~\cite{Hoffmann}).}
    \label{fluo}
\end{figure}

Emission by a fluorescent lamp extends over a finite
interval $[\lambda_{min},\lambda_{max}]$ with $\lambda_{min}$=380 nm and $\lambda_{max}$=700 nm
for this case. The evaluation of the PER is done by spline interpolating the data 
displayed in fig.~\ref{fluo}. Using the general definition eq.~\ref{PER_TB}
we obtain a PER of 343 lm/Watt which represents an efficiency of 50\%.
Usually the conversion efficacy $\eta_C$ is about 20\% which makes the
overall efficacy of 68.6 lm/Watt and the total efficiency at 10 \%.
Such efficiency is quite interesting, however the problem with fluorescent light is
that it suffers from flicker (fluctuating light intensity)
due to the ballast and the random collision phenomena, besides
it relies on Mercury which is a highly polluting source of the environment.
Moreover the ballast circuit might produce annoying noise in some cases.

\end{enumerate}

\subsection{Luminous efficacy of lasers and LED's}

White light, Black-Body and fluorescent radiators are considered as broadband emitters since
their radiation spans (at least) the entire visible spectrum.

This is not the case of LED's and lasers since they are somehow closer to
monochromatic (narrowband) sources. 

The new CIE definition of the mechanical equivalent of the lumen
is that a monochromatic source that is a power emission spectrum peaking at
$\lambda_0$=555 nm (at the maximum sensitivity of the eye) 
with power of 1 W produces exactly 683 lumens.

This can be understood readily from the general PER definition:

\be
\eta_P=K_m \frac{  \int_{\lambda_{min}}^{\lambda_{max}} \delta(\lambda-\lambda_0) V(\lambda) d\lambda}
{ \int_{\lambda_{min}}^{\lambda_{max}} \delta(\lambda-\lambda_0) d\lambda }
\label{PER_laser}
\ee

where we used: $P(\lambda)=a \delta(\lambda-\lambda_0)$ with $a$ a constant
and the condition that $\lambda_0 \in [\lambda_{min}, \lambda_{max}]$.

The integral gives the result: PER= $K_m V(\lambda_0)= K_m$ since 
$V(\lambda_0=555 \mbox{nm})=1$.

As an application, consider a laser emitting at $\lambda_0$=570 nm with a power of 50 mW.
It has a PER=$K_m V(\lambda_0)$ that produces 30 lumens.

Turning to lighting with LED's, one of the main advantages of LED 
is that its lifetime is extremely long (on the order of 100,000 hours) because it is a rugged solid state device. Moreover it does not
rely on a ballast or Mercury which makes it safer than Fluorescent lamps whose lifetime is
about several 1000 hours.

The LED power emission spectrum function is usually approximated by a Gaussian or
a superposition of several Gaussian functions. 
In the single Gaussian approximation 
$P(\lambda)= \exp \left[-\frac{{(\lambda-\lambda_0)}^2}{2 \Delta \lambda^2}\right]$
can be used to evaluate the efficacy $\eta_P$ from eq.~\ref{PER_TB}.

The LED is characterized by an average wavelength $\lambda_0$ and a standard deviation $\Delta \lambda$.
As an example, we consider a blue LED with $\lambda_0$=450 nm and $\Delta \lambda$=20 nm. The PER
obtained from eq.~\ref{PER_TB} is about 39.7 lm/Watt and therefore an efficiency of 6\%. The small efficiency is due to small overlap between the blue LED spectrum
and the eye sensibility curve $V(\lambda)$, moreover that number is further reduced
after multiplication by the conversion efficacy $\eta_C$ which is typically about 20\%
yielding a total efficiency of 1.2\%. 

This is to be contrasted with the present status of White LED who has a very large PER as illustrated by
Haitz law in fig.~\ref{Haitz}. That might be due to the fact a flat power 
emission spectrum 
enhances the PER as previously seen with White sources however the question might be asked more generally
in specific terms as explained in the next section.

\section{Maximum Luminous efficacy of radiation sources}
An important question can now be asked: For a given color (chromaticity),
is there a maximum PER that can be realized with any radiation source?

Mathematically this can be answered with the following set of assumptions:
Given a source endowed with a normalized power emission spectrum function $P(\lambda)$:

\be
\int_{\lambda_{min}}^{\lambda_{max}} P(\lambda) d\lambda=1
\label{norm}
\ee 

Is it possible to find the best $P(\lambda)$ such that the PER given by:

\be
Y=K_m \int_{\lambda_{min}}^{\lambda_{max}} P(\lambda) \bar{y}(\lambda) d\lambda
\ee

for some given color represented by chromaticity coordinates $x_c,y_c$ (see Appendix)
is maximized. Note that this stems from the fact  $\bar{y}(\lambda)= V(\lambda)$
as explained in the Appendix.

Thus we have an optimization problem for an unknown function $P(\lambda)$
subject to three constraints: normalization eq.~\ref{norm} and fixed color ($x_c,y_c$) 
constraints eq.~\ref{XY}.

Following Ohta \etal   suggestion~\cite{Ohta}, we transform the problem into its discrete
version by dividing the wavelength interval $[\lambda_{min}, \lambda_{max}]$ into $N$ values 
$\lambda_{i}$ with a step $\Delta \lambda$ such that the objective function to be optimized is:

\be 
\mbox{max} \int P(\lambda) \bar{y}(\lambda) d\lambda \rightarrow \mbox{max} \sum_{i=1}^N P_i \bar{y}_i 
\label{optimum}
\ee

Discrete values $P_i, \bar{x}_i , \bar{y}_i, \bar{z}_i$ correspond to
$ P(\lambda), \bar{x}(\lambda), \bar{y}(\lambda), \bar{z}(\lambda)$ taken at
$\lambda= \lambda_{i}$. 

The normalization constraint eq.~\ref{norm} becomes:

\be 
\Delta \lambda \sum_{i=1}^N P_i=1 
\label{cons1}
\ee

whereas the chromaticity constraints (see Appendix) become:

\bea
x_c= \frac{ \sum_{i=1}^N P_i \bar{y}_i }{ \sum_{i=1}^N P_i (\bar{x}_i+\bar{y}_i+\bar{z}_i)} \nonumber \\
y_c= \frac{ \sum_{i=1}^N P_i \bar{y}_i }{ \sum_{i=1}^N P_i (\bar{x}_i+\bar{y}_i+\bar{z}_i)}
\eea

This can be transformed into:

\bea
\sum_{i=1}^N P_i [ x_c (\bar{x}_i+\bar{y}_i+\bar{z}_i) -\bar{x}_i]=0, \nonumber \\
\sum_{i=1}^N P_i [ y_c (\bar{x}_i+\bar{y}_i+\bar{z}_i) -\bar{y}_i]=0
\label{cons2}
\eea

The problem now is expressed in the standard Simplex form (see Numerical Recipes~\cite{NR}, chapter 10):
One ought to find a set of $N$ values $P_i$ such that eq.~\ref{optimum} is maximized
under three constraints given by eq.~\ref{cons1} and eqs.~\ref{cons2}.

The Simplex method~\cite{NR} results are displayed in
fig.~\ref{MAX_PER}. The constant PER curves or iso-PER curves represented in the CIE
diagram (see Appendix) get closer to the CIE contour as the PER is increased.

Low values of PER are in the blue region of the CIE diagram which explains 
the result obtained for the blue diode, whereas larger PER values occur 
as we move toward the yellow part of the CIE diagram.

We find the largest value of 679 lm/Watt, as before, and the corresponding 
iso-PER curve in the vicinity of the 555 nm region, the area of highest 
sensibility of the eye, confirming the candela standard once again 
and the SI metrological data~\cite{BIPM}.

\begin{figure}[htbp]
  \centering
    \includegraphics[angle=0,width=80mm,clip=]{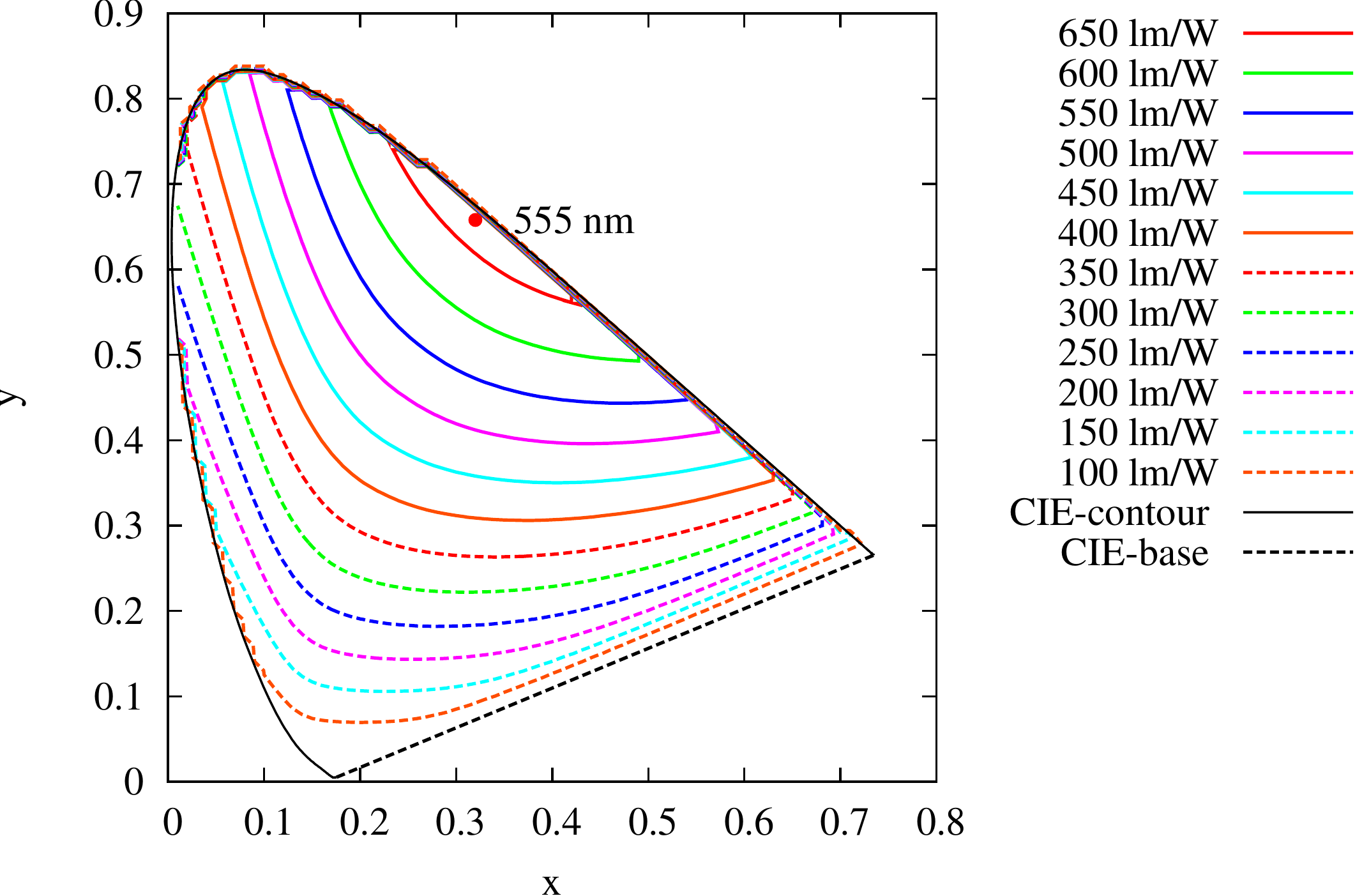}  
\caption{(Color online) Constant PER curves in the CIE diagram. 
The maximum value of 679 lm/Watt is
attained around the 555 nm region (red spot) as adopted by the SI system. Waviness
observed in some curves around borderline are due to slow convergence of the 
optimization simplex algorithm.}
    \label{MAX_PER}
\end{figure}

\section{Discussion and Conclusion}
Some lighting metrics have been introduced and perspectives for future development
regarding the increase of lighting intensity and quality were presented. \\
Despite the fact White LED is presently showing tremendous potential in terms
of quality and PER increase, the diagram presented in fig.~\ref{MAX_PER} indicates
that a white source maximum is between 350 and 400 lm/Watt
which is yet to be reached by White LED's. \\
On the other hand, Yellow-Green light sources may achieve a large increase since their
maximum PER is around the theoretical standard of 683 lm/Watt adopted by both
CIE and SI organizations. \\
The conclusion is that work must be targeted toward increasing rather
the conversion efficacy $\eta_C$ in the White sources case and in particular
the White LED case. 

\appendix

\section{CIE Chromaticity Coordinates}

In 1931, the CIE undertook a series of historical measurements called Color 
Matching experiments in order to calibrate colorimetry and 
human color perception.
A number of "standard" observers had to make a comparison between
a color of a given wavelength $\lambda$ and 
a superposition of three selected wavelengths called RGB primaries~\cite{primaries}. 
The weight of each of the three colors to perform the match was recorded.
The observations were done at a fixed distance of 50 cm with two possibilities for the 
eye angle opening (defined by the observation diameter value) of 2\deg
and 10\deg. 

This led to the existence of color matching functions (CMF) $\bar{r}(\lambda), 
\bar{g}(\lambda),\bar{b}(\lambda)$ corresponding to the red, green, blue
weight coefficients that matched the color $\lambda$.

The study showed that not all matching weights are positive and that some
values were negative.
The algebraic values of the coefficients meaning that CMF
functions took positive and negative values originating from the overlap 
versus wavelength between human cone sensitivities. 

\begin{figure}[htbp]
  \centering
    \includegraphics[angle=0,width=80mm,clip=]{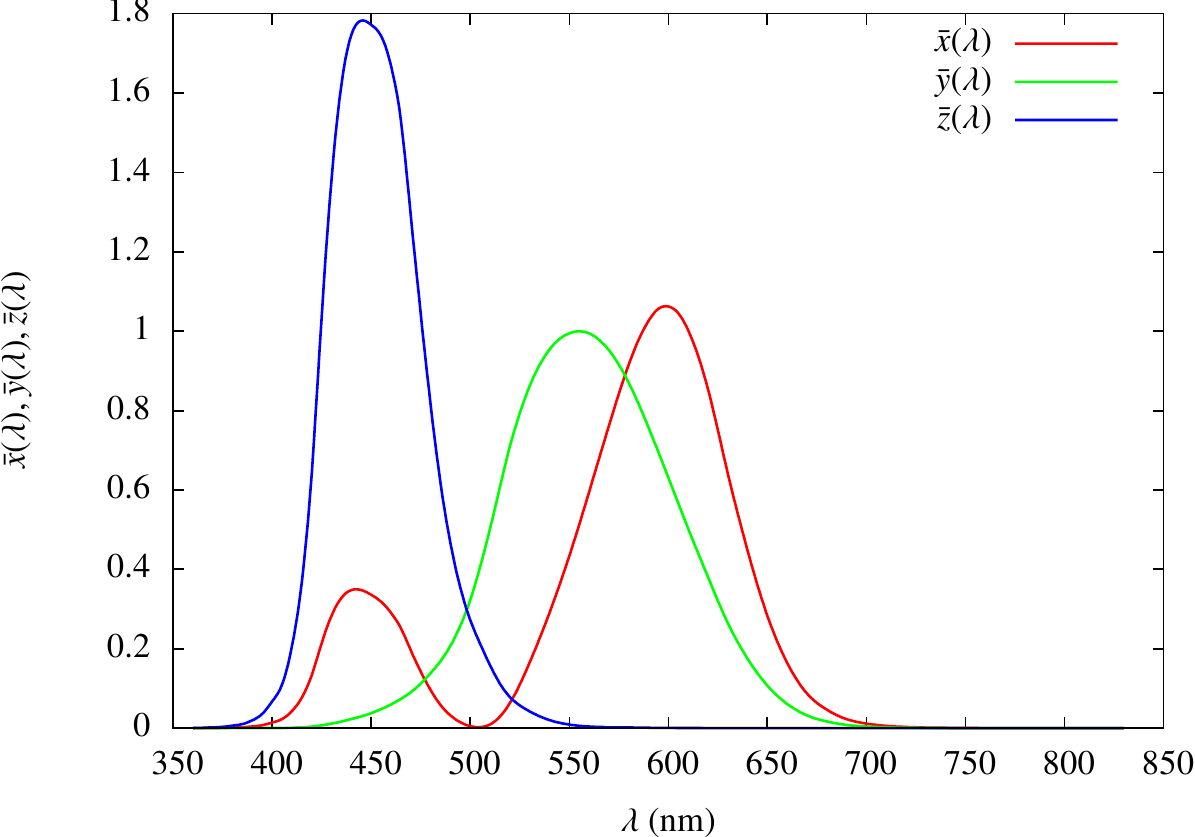}  
\caption{(Color online) Color matching functions of the CIE for eye opening of 2\deg.
$\bar{x}$ in red, $\bar{y}$ in green and $\bar{z}$ in blue cover approximately
the corresponding RGB color zones.}
    \label{tristimulus}
\end{figure}

This pushed the CIE
to perform a linear transformation over $\bar{r}(\lambda), 
\bar{g}(\lambda),\bar{b}(\lambda)$ in order to define three strictly
positive CMF functions $\bar{x}(\lambda), 
\bar{y}(\lambda),\bar{z}(\lambda)$ displayed in fig.~\ref{tristimulus}. 

The linear transformation is based on equal area of $\bar{x}(\lambda), 
\bar{y}(\lambda),\bar{z}(\lambda)$ over the visible spectrum and the
choice for the middle spectrum function $\bar{y}(\lambda)$ to be taken
equal to $V(\lambda)$, the photopic eye sensitivity.

If we have a radiation source characterized by a power emission spectrum function $P(\lambda)$
its tristimulus coordinates are given by:

\bea
X=K_m \int_{\lambda_{min}}^{\lambda_{max}} P(\lambda) \bar{x}(\lambda) d\lambda, \nonumber \\ 
Y=K_m \int_{\lambda_{min}}^{\lambda_{max}} P(\lambda) \bar{y}(\lambda) d\lambda, \nonumber \\ 
Z=K_m \int_{\lambda_{min}}^{\lambda_{max}} P(\lambda) \bar{z}(\lambda) d\lambda, \nonumber \\
\label{chroma}
\eea

\begin{figure}[htbp]
  \centering
    \includegraphics[angle=0,width=80mm,clip=]{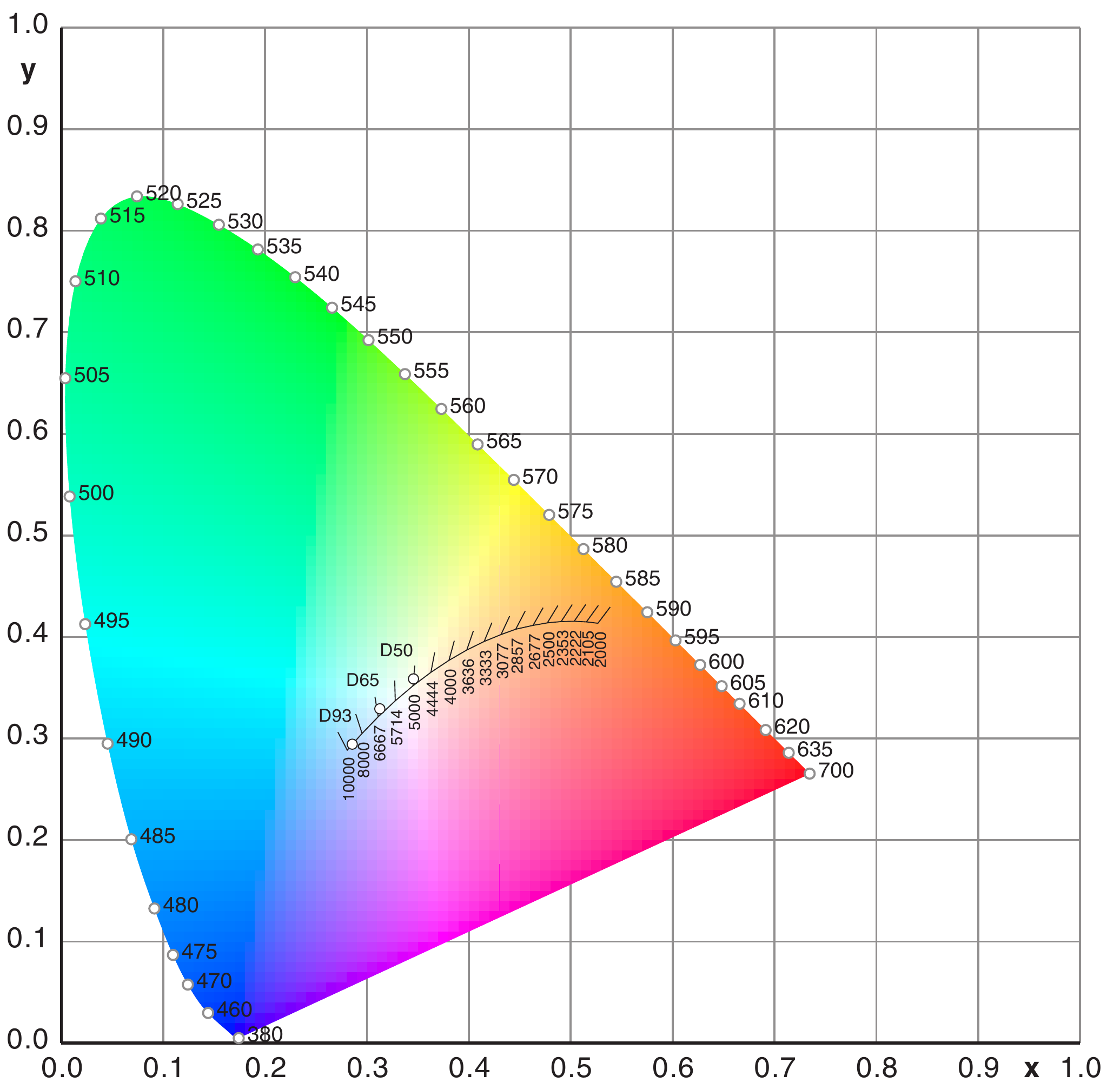}  
\caption{(Color online) CIE diagram displaying color of points with
coordinates $(x,y)$ and the Black-Body radiation color path
as a function of absolute temperature. The various symbols D$_T$ correspond to
Daylight type source (illuminant or synthetic source) at a given temperature $T/100$. 
For instance D$_{65}$ is for T=6500K (adapted from Hoffmann~\cite{Hoffmann}).}
\label{CIE}
\end{figure}  

The color of the  $P(\lambda)$ source is given by a point with coordinates $(x,y)$
in the CIE diagram displayed in fig.~\ref{CIE}. $(x,y)$ are called chromaticity 
coordinates with values explicitly given by:

\be
x=\frac{X}{X+Y+Z}, y=\frac{Y}{X+Y+Z}
\label{XY}
\ee

The CIE diagram (called tongue or horseshoe diagram) shown in fig.~\ref{CIE} 
displays several interesting characteristics:
\begin{enumerate}
\item The contour contains pure colors (completely saturated or free of any white content)
having wavelengths indicated on the borderline in nanometers. 
The corresponding wavelengths are called dominant since they control the color
from pure (on the border) to White point at the center with coordinates
$x=\frac{1}{3}, y=\frac{1}{3}$.
\item Colors within the horseshoe diagram are unsaturated and as we move forward
toward the White point they become pastel like. This stems from the increase of white
content as we proceed toward the white point.
\item Black-Body color appears on a path as a function of absolute temperature. 
It follows the red, orange, yellow, white and finally blue sequence as temperature
is increased. This describes heated metals and agrees with Kirchhoff observations~\cite{Grum}.
\item Illuminants (artificial daylight sources) indicated by D$_{50}$, D$_{65}$ and D$_{93}$ 
appear at their corresponding colour with index (50,65,93) equal to 
absolute temperature divided by 100. Black-Body sources with temperatures
of 5000K, 6500K and 9300K have colours close to the White point.
\item The CIE contour is closed from below by a straight line (called also
the purple line) that does not carry any dominant wavelength. It means that
most purple colors cannot be obtained by altering the White content
of some main (dominant) color as done before. This is another consequence
of the cone overlap that resulted in negative CMF weights.  
 
\end{enumerate}

\end{document}